# Theoretical Evaluation of the Capacity-Achieving Distribution for IM-DD Fiber-Optic Channels

Dongdong Zou, Wei Wang, Sui Qi, Fan Li, and Zhaohui Li

*Abstract*— **The capacity and capacity-achieving distribution for intensity-modulation and direct-detection (IM-DD) fiber-optic channels is theoretically investigated. Different from coherent fiber-optic channels, we indicate that the capacity-achieving distribution of IM-DD systems should be discussed separately in two cases: 1) IM-DD systems without optical amplifier, which are constrained in peak power; 2) IM-DD systems with optical amplifier, which are the average power constraint (APC) system. For the two models, the maximum mutual information achieving distribution, instead of the maximum input entropy achieving distribution, is numerically computed by the iterative Blahut-Arimoto (BA) algorithm. For the IM-DD system under peak power constraint (PPC), a dynamic-assignment BA algorithm is applied to find the capacity-achieving distribution with minimum cardinality. It is observed that the maximum difference between the minimum input cardinality and capacity is around 0.8 bits. For a fixed support input cardinality, although the observed shaping gain is small and only appears in low peak-signal-to-noise ratio (PSNR) regions in the PPC IM-DD system, the probabilistic shaping technique can also be used to introduce rate adaptation to the system by adjusting the shaping and FEC overheads since the capacity-achieving distribution is symmetric. In the IM-DD system under APC, a modified BA algorithm is investigated to solve for the capacity and capacity-achieving distribution, and a significant shaping gain is observed. For PAM8 and PAM16 modulation formats, 0.294 bits/symbol and 0.531 bits/symbol shaping gain can be obtained at the SNR of 20dB. Furthermore, since the capacity-achieving distribution is asymmetric in this case, a practical discussion of the PS technique is also presented.**

## I. INTRODUCTION

**I**NTENSITY modulation and direct-detection (IM-DD) have been widely applied in numerous scenarios such as free space optical (FSO) links, indoor visible light communications, optical wireless communications, and especially short-reach fiber-optic links due to the low system cost and complexity [1-4]. With the continuous deployment of data centers, more energy-efficient and flexible short-reach optical datacenter interconnects (DCI) are highly desirable, which arouses extensive research on probabilistic shaping (PS) technique in IM-DD optical fiber communication systems [5-16].

Actually, PS debuted in long-haul coherent optical fiber communication systems and achieved remarkable success in 2015 [17]. The well-known 1.53 dB shaping gain in coherent optical fiber communication systems benefits from two facts: (i) optical amplifiers are required to boost the transmitter side signal to yield the desired launch power and compensate for the fiber loss over a long-haul transmission, resulting in an average power constraint (APC) model with additive white Gaussian noise (AWGN) features; (ii) in an AWGN channel with APC, Maxwell-Boltzmann (MB) signaling achieves the maximum information rate. Furthermore, by applying the constant composition distribution matcher (CCDM) [18] and forward error correction (FEC) encoder in parallel, namely probabilistic amplitude shaping (PAS) structure, the fine-granularity rate adaption can be easily realized by changing the rate of the CCDM while with a constant FEC code rate [19]. Due to the superiority of the PS technique, it has been widely investigated for IM-DD systems. In [5-10], the application of PS technique is exactly the same as coherent systems, in which the probability of the discrete pulse amplitude modulation (PAM) symbols is shaped into the Gaussian distribution. Compared to long-haul coherent systems, IM-DD systems exhibit some unique characteristics: (i) signal is linearly modulated from the electric field to optical intensity rather than optical field; (ii) the information is only conveyed by the intensity of the optical carrier, thus there is a nonnegative constraint on the input signal; (iii) optical amplifier is not an essential device in IM-DD systems, which means that the system is limited by peak power instead of average power in the IM-DD system without optical

Manuscript received XXX XXXX; revised XXX, XXXX; accepted XXX, XXXX. This work is partly supported by the National Key R&D Program of China (2018YFB1801704); the National Natural Science Foundation of China (U2001601, 61871408); Local Innovation and Research Teams Project of Guangdong Pearl River Talents Program (2017BT01X121); Fundamental and Applied Basic Research Project of Guangzhou City under Grant (202002030326); Open Fund of IPOC (BUPT) (IPOC2020A010). (*Corresponding Author: Fan Li*)

D. Zou, W. Wang, and F. Li are with the School of Electronics and Information Technology, Guangdong Provincial Key Laboratory of Optoelectronic Information Processing Chips and Systems, Sun Yat-Sen University, Guangzhou 510275, China (e-mail: lifan39@mail.sysu.edu.cn).

Q. Sui is with Southern Marine Science and Engineering Guangdong Laboratory (Zhuhai), Zhuhai 519080, China (e-mail: suiqi@sml-zhuhai.cn).

Z. Li is with School of Electronics and Information Technology, Guangdong Provincial Key Laboratory of Optoelectronic Information Processing Chips and Systems, Sun Yat-Sen University, Guangzhou 510275, China, and he is also with Southern Marine Science and Engineering Guangdong Laboratory (Zhuhai), Zhuhai 519080, China (e-mail: lzhh88@mail.sysu.edu.cn).

amplifier. These features imply that the MB/Gaussian signaling is largely not the capacity-achieving distribution of IM-DD systems. For IM-DD systems with optical amplifier, such as intra-DCI (40~80km), the average power is represented by the first moment rather than the second moment of the transmitted signal due to the characteristics (i) and (ii) mentioned above. Thus, in [11-13, 20], the exponential distribution is treated as the capacity-achieving distribution. However, the exponential distribution is only the maximum entropy achieving solution under the constraint of the first moment, and the system capacity is defined as the maximum achieved mutual information between the transmitted and received signal. The shaping gain gap between the capacity-approaching distribution and the exponential distribution remains an open question in IM-DD systems under APC. For IM-DD systems under peak power constraint (PPC), such as the intra-DCI without the optical amplifier, different schemes have been evaluated [14-16]. In [14], the authors find the optimal solution for the PAM4 modulation format by performing a linear search over a single parameter $P_X(0)$, since the optimal distribution is symmetric around the mean value $E(X)$. However, this optimization approach cannot be directly applied to higher modulation formats. In [15], the authors give out an optimal probability mass function (PMF) for the PAM6 symbol but without any explanation. In [16], Gaussian signaling is compared with uniform signaling, and the authors conclude that Gaussian signaling can outperform uniform signaling when the system bandwidth limitation is serious. However, there is no evidence that Gaussian distribution is the capacity-achieving distribution in the IM-DD system under PPC.

Although a number of works have been reported on the application of PS technique in IM-DD fiber-optic systems, a systematic and theoretical evaluation of the capacity-achieving distribution in this system has not been involved. In this paper, we focus on the discussion of the capacity-achieving distribution for the IM-DD systems without or with slight bandwidth limitation, since only a low complexity feedforward equalizer (FFE) with few taps is allowed at the receiver in a practical short-reach communication system.

The main contributions of this paper are as follows:

- We point out that when seeking the capacity and capacity-achieving distribution of the IM-DD optical communication system, two models, namely the peak power constraint and average power constraint systems, should be discussed separately, which are related to the IM-DD system without and with optical amplifier, respectively. For the two system models, the capacity-achieving PMFs are distinct and quite different from the MB/Gaussian distribution.
- Unlike the coherent communication system, there is no analytic solution for the capacity-achieving distribution of the IM-DD system. We propose to utilize the modified Blahut-Arimoto (BA) algorithm to iteratively solve for the capacity and capacity-achieving distribution for the mentioned two IM-DD systems.
- For IM-DD systems under PPC, a dynamic-assignment Blahut-Arimoto (DAB) algorithm is applied to find the minimum cardinality capacity-achieving distribution. It is observed that the maximum difference between the minimum input cardinality and capacity is around 0.8 bits. For the fixed support constellation cardinality, the capacity-achieving distribution is numerically solved by the traditional BA algorithm. We find that the optimal PMF of the transmitted signal exhibits a cup-shaped symmetric distribution, rather than the uniform or Gaussian distribution. Although the observed shaping gain is small and can only be obtained in low peak-signal-to-noise ratio (PSNR) regions, the PS technique can also be utilized to introduce rate adaptation to the system, by adjusting the shaping and FEC overheads.
- For the IM-DD system under APC, a modified BA algorithm is investigated to numerically evaluate the capacity and capacity-achieving distribution, and a significant shaping gain is observed. For PAM8 and PAM16 modulation formats, 0.294 bits/symbol and 0.531 bits/symbol shaping gain can be obtained at the signal-to-noise ratio (SNR) of 20dB. In addition, the shaping gain between our solved capacity-achieving distribution and exponential distribution is compared. The maximum shaping gain gap for the PAM8 format is lower than 0.05 bits/symbol.
- The practicality and rate adaptability of the proposed PS scheme in the two IM-DD optical communication systems are also discussed. For the IM-DD system under PPC, the BA solved optimal PMF shows a symmetric distribution, thus it is very easy to adopt the PAS structure to achieve the rate adaption by adjusting the overhead of CCDM and FEC. For the IM-DD system under APC, the capacity-achieving distribution is asymmetric. Thus, a modified exponential distribution with twin probability shape is investigated to realize PAS.

The remainder of this paper is organized as follows. Section II develops the system models for short-reach fiber-optic channels. Section III gives out the method for the capacity and capacity-achieving distribution evaluation. In Section IV we present numerical results, and in Section V we conclude the paper.

## II. SYSTEM MODEL

In information theory, a basic issue is to find the capacity and capacity-approaching input distribution for a given channel model [21]. The most well-studied channel model is the discrete memory-less AWGN channel

$$Y_k = X_k + Z_k, \ k = 1, 2, \ldots \quad (1)$$

where $Y_k$ is the output of a channel, $X_k$ and $Z_k$ are the independent channel input and additive white Gaussian noise, respectively. In this model, the channel can be characterized by the conditional probability density function

$$p(y \mid x) = \frac{1}{\sqrt{2\pi\sigma^2}} \exp(-\frac{\| y - x \|^2}{2\sigma^2}). \quad (2)$$

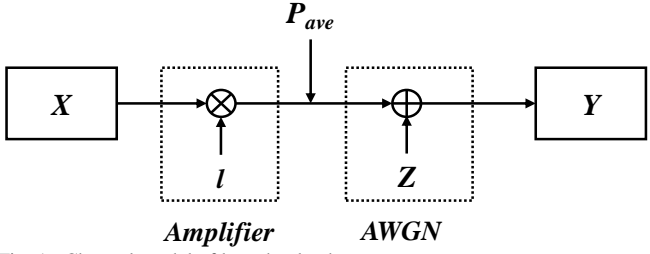

Fig. 1. Channel model of long-haul coherent systems

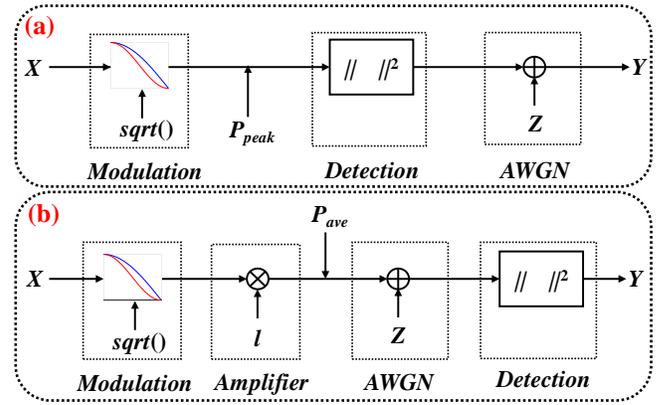

Fig. 2. Channel model of IM-DD systems (a) without optical amplifier (b) with optical amplifier.

where $\sigma^2$ is the variance of the white Gaussian noise. The average mutual information (MI) can be expressed as [21]

$$I(X;Y) = E_{X,Y}\left\{\log_2\left[\frac{p(y \mid x)}{p(y)}\right]\right\} \qquad (3)$$

where $p(y)$ is the probability density function of the received signal. The channel capacity is given by

$$C = \max_{p(x)} I(X;Y). \qquad (5)$$

where $p(x)$ is the PMF of the input signal. Overall probability distributions of the input signal, channel capacity is the maximum MI achieved. In a practical system, different constraints such as PPC, APC, and fixed finite support constellation size will be imposed. Thus, solving the capacity and capacity-achieving distribution is an optimization problem in a finite field. In this paper, we focus on the discussion of the fiber-optic channels including coherent and IM-DD systems without or with slight bandwidth limitation. We briefly review the channel model and the capacity-achieving distribution of long-haul coherent systems, and then the target system, namely IM-DD systems, will be discussed in detail.

### A. Channel model of Coherent Systems

For long-haul coherent systems, the channel can be well characterized by the AWGN model described in Eq. (1). In such a system as shown in Figure 1, the optical amplifier is required to boost the transmitted signal to yield the desired launch power or compensate the transmission loss, which results in an average power constraint $P_{ave}$

$$E[\|l \cdot X\|^2] \le P_{ave}. \qquad (5)$$

where $E[.]$ is the mean operation, $l$ is a scaling factor to model amplification of the optical amplifier. In this system, the noise $Z$ is dominated by the amplified spontaneous emission noise (ASE). Considering the input symbol is constrained to a finite constellation size $\mid \chi \mid = M$

$$\chi = \{x_1, x_2, ..., x_M\}. \qquad (6)$$

the PMF of the input symbol can be expressed by

$$p_\chi = [p_{x_1}, p_{x_2}, ..., p_{x_M}]. \qquad (7)$$

Thus, the target is to find the capacity $C$ and capacity-achieving input distribution $p_\chi^*$

$$C = \max_{P_\chi^*, l} I(X;Y). \qquad (8)$$

$$p_{\chi^*} = [p_{x_1}^*, p_{x_2}^*, ..., p_{x_M}^*]. \qquad (9)$$

under the constraints

$$\sum_{k=1}^{M} p_{x_i}^* = 1. \qquad (10)$$

$$\sum_{k=1}^{M} p_{x_i}^* \cdot \| l \cdot x_i \|^2 \le P_{ave}. \qquad (11)$$

For a more specific expression, the MI can be further given by

$$I(X;Y) = H(Y) - H(Y \mid X) = H(Y) - H(Z). \qquad (12)$$

where $H(Y)$ and $H(Z)$ are the entropy of received signal and noise, respectively. For a Gaussian white noise with variation $\sigma^2$, the entropy is a constant

$$H(Z) = \frac{1}{2}\log_2(2\pi e\sigma^2). \qquad (13)$$

Thus, the issue is transformed to find the maximum entropy of the received continuous signal

$$\max_{p(y)} H(Y) = -\int p(y)\log_2[(p(y)]dy. \qquad (14)$$

under the constraints

$$\int p(y)dy = 1. \qquad (15)$$

$$\int p(y)\| y \|^2 \, dy \le P_{ave}^{'}. \qquad (16)$$

where $P_{ave}^{'}=P_{ave}+\sigma^2$ is the power of received signal. By utilizing the Lagrange multiplier method, the cost function can be expressed as

$$J = H(Y) + \lambda\left[\int p(y)dy - 1\right] + \mu\left[\int p(y)\| y \|^2 dy - P_{ave}^{'}\right]. \qquad (17)$$

where $\lambda$ and $\mu$ are Lagrange multipliers. Consequently,

$$p^*(y) = \frac{1}{\sqrt{2\pi P_{ave}^{'}}}\exp(-\frac{y^2}{2P_{ave}^{'}}). \qquad (18)$$

Since both the receiver side signal $Y$ and the noise $Z$ satisfy Gaussian distribution, the optimal PMF of the input signal is the Gaussian distribution.

$$p_{x_i}^* = \frac{1}{\sqrt{2\pi P_{ave}}}\exp(-\frac{\| x_i \|^2}{2P_{ave}}), k = 1, 2, ..., M. \qquad (19)$$

As for the shaping gain, it can be expressed as the difference between the entropy of the received signal with the Gaussian distribution input and uniform input according to Eq. (12) [21]

$$\Delta H = H(Y_G) - H(Y_U) \le H(X_G) - H(X_u). \qquad (20)$$

$$H(X_G) = \frac{1}{2}\log_2(2\pi eP_{ave}). \qquad (21)$$

$$H(X_U) = \frac{1}{2}\log_2(12P_{ave}). \qquad (22)$$

when SNR goes to infinity, the shaping gain in power efficiency

is given by

$$\Delta P_{dB} = \frac{\partial P_{dB}}{\partial P} \cdot \frac{\partial P}{\partial H} \cdot \Delta H = 1.5329 dB. \tag{23}$$

### B. Channel model of IM-DD Systems

As mentioned in Section I, the channel model of IM-DD systems is quite different from coherent systems, which implies that the capacity-approaching distribution is no longer the Gaussian distribution. To exact evaluation of the capacity-achieving distribution for IM-DD systems, we further divide the IM-DD systems into the system without and with optical amplifier as shown in Figs. 2(a) and 2(b), respectively.

#### 1) IM-DD systems without optical amplifier

For a typical short-reach IM-DD optical fiber communication system such as intra datacenter interconnects (intra-DCI, <2km), the optical amplifier is not equipped due to the system cost limitation and low transmission loss. In such a system as shown in Fig. 2(a), the nonnegative transmitted sequence $X$ is nonlinearly modulated into the optical domain, namely linear modulation from the electric field to optical intensity. We denote the optical field signal by

$$X^{'} = \sqrt{X}. \tag{24}$$

As no optical amplifier is applied, the PPC is valid

$$\max(\|X^{'}\|^2) \le P_{peak}. \tag{25}$$

At the receiver side, the signal is detected by a photodetector (PD). In this channel model, the system performance is predominantly determined by the receiver side white Gaussian noise, including the thermal noise of PD and radio frequency (RF) amplifiers such as transimpedance amplifier (TIA) [16]. Therefore, we only model a white Gaussian noise after the optical-to-electrical (O-E) conversion at the receiver. Finally, the IM-DD system without optical amplifier can be modeled as a discrete memoryless channel

$$Y_k = X_k + Z_k. \tag{26}$$

The conditional probability density function is given by

$$p(y \mid x) = \frac{1}{\sqrt{2\pi\sigma^2}} \exp(-\frac{\|y - x\|^2}{2\sigma^2}). \tag{27}$$

We are interested to calculate the capacity $C$ and capacity-approaching input distribution $P_X^*$

$$C = \max_{p_X} \left[ I(X;Y) \right]. \tag{28}$$

Under the constraints

$$x_k \ge 0, k = 1, 2, ..., M. \tag{29}$$

$$\sum_{k=1}^{M} p_{x_k}^* = 1. \tag{30}$$

$$\max[x_k] \le P_{peak}, k = 1, 2, ..., M. \tag{31}$$

Unlike long-haul coherent systems, the capacity-achieving distribution for IM-DD systems with PPC cannot be directly solved by Lagrange multiplier method since the transmitted signal is unipolar but the noise is bipolar. However, when the system SNR goes to infinity, the system capacity-achieving distribution tends to the maximum input entropy achieving distribution. Now we utilize the Lagrange multiplier method to find the maximum input entropy achieving distribution under the constraints of Eqs. (29)-(31).

$$H^* = \max_{p_X} H(X). \tag{32}$$

The cost function is

$$J = H(X) + \lambda(\sum_{k=1}^{M} p_{x_k} - 1) + \mu[\max(x_k) - P_{peak}]. \tag{33}$$

In Appendix, we show that the solution is the uniform distribution. We draw a conclusion that the uniform distribution is the local optimal distribution for IM-DD systems with PPC and extremely high SNR. Consequently, when SNR approaches infinity, there is no shaping gain in this system. Although several distributions have been investigated in this system for shaping gain [10, 14-16], none of them have been proved from the perspective of information theory. In Section III, a theoretical evaluation of the capacity-achieving distribution for the IM-DD system under PPC is presented by an iterative method. Furthermore, the proposed method can be utilized to solve for the optimal input distribution in a wide range of SNR.

#### 2) IM-DD systems with optical amplifier

IM-DD structure is not only a preferred choice for short-reach applications such as intra-DCI, but an attractive solution for longer distance transmission such as inter-DCI (40~80km) due to the merits of low complexity and cost. For this kind of application, O-band transmission is the most commonly utilized scheme due to its robustness against chromatic dispersion. However, it exhibits higher transmission loss, resulting in the requirement of optical amplifier to boost the signal to the desired launch power or compensate for the transmission loss, which invalidates the PPC and imposes the APC in this system. The channel model of this kind of system is shown in Fig. 2(b). As optical ASE is the predominant noise, an AWGN channel is modeled in the optical domain before the direct detection receiver. The discrete-time model of the channel should be defined as

$$Y_k = |\sqrt{X_k} + Z_k|^2. \tag{34}$$

It can be found that the channel is different from the two systems mentioned above. The conditional probability density function is a non-central chi-square distribution with one degree of freedom [20, 22, 23] rather than Gaussian distribution

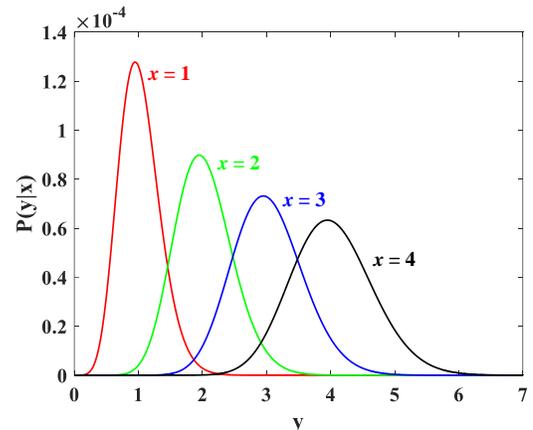

Fig. 3. $p(y|x)$ of APC IM-DD system with PAM4 format and 20dB SNR.

$$p(y \mid x) = \frac{1}{2\sigma^2}(\frac{y}{x})^{-\frac{1}{4}}\exp(-\frac{y+x}{2\sigma^2})I_{-\frac{1}{2}}\left[\frac{\sqrt{yx}}{\sigma^2}\right]. \quad (35)$$

where $I()$ is the modified Bessel function of the first kind, $y$ and $x$ are the nonnegative intensity of received and transmitted signal, respectively. Due to the square-law detection, the system performance is dominated by the signal-ASE beating noise. When the ASE-ASE beating noise is omitted, conditioned on the intensity $x_k$, the received signal $y$ shows an approximate Gaussian distribution with mean $x_k$, and variance

$$\sigma_k^2 = 2x_k\sigma^2. \quad (36)$$

For example, we give out the $p(y|x=k)$ of the PAM4 modulation format with 20dB SNR in Fig. 3 (for easy observation, $X$ with [1 2 3 4] rather than [0 1 2 3] is utilized). Clearly, in such a high SNR, the $p(y|x)$ for a given $x_k$ is closed to Gaussian distribution, and the variance is proportional to the signal intensity $x_k$. For this system, we want to solve for the capacity $C$ and capacity-achieving distribution $P_X^*$ under the constraints

$$x_k \geq 0, k = 1, 2, ..., M. \quad (37)$$

$$\sum_{k=1}^{M} p_{x_k}^* = 1. \quad (38)$$

$$\sum_{k=1}^{M}(p_{x_k}^* \cdot l \cdot x_k) \leq P_{ave}. \quad (39)$$

As mentioned in the previous subsection, an analytic solution cannot be directly found for IM-DD systems in a wide range of SNR. We can utilize the Lagrange multiplier method to solve the local optimal input distribution for the infinite SNR case. The cost function can be expressed as

$$J = H(X) + \lambda(\sum_{k=1}^{M} p_{x_k} - 1) + \mu\left[\sum_{k=1}^{M}(p_{x_k} \cdot l \cdot x_k) - P_{ave}\right]. \quad (40)$$

The solution for this issue is the exponential distribution which has been discussed in Appendix. The shaping gain can be calculated by

$$\Delta H = H(X_E) - H(X_U). \quad (41)$$

$$H(X_E) = \log_2(eP_{ave}). \quad (42)$$

$$H(X_U) = \log_2(2P_{ave}). \quad (43)$$

Consequently, the shaping gain in power efficiency of 1.3326dB can be obtained. The detailed calculation process is also presented in Appendix. It should be noted that the 1.3326dB is only the shaping gain when SNR goes to infinite, not the upper bound of the shaping gain for this system. Since the exponential distribution is only the local optimal solution (high SNR region) rather than the global optimal solution for the IM-DD systems under APC, the monotonicity of the shaping gain cannot be guaranteed as in the coherent system. In order to find the capacity-achieving distribution of the IM-DD system under APC in a wide range of SNR, a numerical optimization method is presented in Section III. According to the proposed scheme, the difference in shaping gain between capacity-achieving distribution and exponential distribution is numerically calculated.

## III. METHOD

Section II discusses the system model of coherent systems, IM-DD systems without and with optical amplifier. In long-haul coherent fiber-optic systems with a finite support constellation size, the discrete Gaussian distribution is the global optimal capacity-achieving solution. However, for two kinds of IM-DD systems, it is failed to find an analytic global optimal solution for capacity-achieving distribution, and only two local optimal distributions (in ultra-high SNR regions), namely uniform distribution and exponential distribution, are obtained. In this section, we propose to utilize the BA algorithm [24, 25] to iteratively solve for the capacity-achieving distribution in a wide range of SNR regions for the two IM-DD systems.

### A. Blahut-Arimoto algorithm

In 1948, Shannon introduced the concept of channel capacity to specify the maximum information rate which can be reliably conveyed over a channel [21]. Furthermore, an analytic solution for capacity-achieving distribution is theoretically evaluated for some channel models. However, the analytic solution cannot be found in most cases such as the aforementioned two IM-DD systems. As a numerical programming method, the BA algorithm is proposed to iteratively solve for the capacity $C$ and capacity-achieving distribution $p_X^*$ for discrete memoryless channels based on the observation of

$$C = \max_{p_X}\sum_{X}\sum_{Y} p(x) \cdot p(y \mid x) \cdot \log_2\left[\frac{p(x \mid y)}{p(x)}\right]. \quad (44)$$

where $p(x/y)$ is posterior probability function which is determined by the input distribution $p(x)$ and conditional probability density function $p(y/x)$

$$p(x \mid y) = \frac{p(x)p(y \mid x)}{p(y)}. \quad (45)$$

Recalling the Lagrange multiplier method to calculate the capacity $C$ and capacity-achieving distribution $p_X^*$, the cost function can be expressed as

$$J = \sum_{X}\sum_{Y} p(x) \cdot p(y \mid x) \cdot \log_2[\frac{p(x \mid y)}{p(x)}] + \lambda[\sum_{X} p(x) - 1]. \quad (46)$$

$$\frac{\partial J}{\partial p(x)} = \sum_{Y}\left\{p(y \mid x) \cdot \log_2\left[p(x \mid y)\right]\right\} - \log_2\left[p(x)\right] - 1 + \lambda = 0. \quad (47)$$

Consequently,

$$p(x) = \frac{2^{\sum_{Y} p(y \mid x) \cdot \log_2\left[p(x \mid y)\right]}}{\sum_{X}\left\{2^{\sum_{Y} p(y \mid x) \cdot \log_2\left[p(x \mid y)\right]}\right\}}. \quad (48)$$

Substituting Eq. (45) into Eq. (48)

$$p(x) = \frac{p(x) \cdot 2^{D\left[p(y \mid x) \| p(y)\right]}}{\sum_{X}\left\{p(x) \cdot 2^{D\left[p(y \mid x) \| p(y)\right]}\right\}}. \quad (49)$$

$$D\big[p(y\,|\,x)\,\|\,p(y)\big]=\sum_{Y}p(y\,|\,x)\log_2\left[\frac{p(y\,|\,x)}{p(y)}\right]. \quad (50)$$

The form of Eq. (49) indicates that the capacity-achieving distribution $p(x)$ shows a recursive feature, namely a new $p(x)$ in the left can be generated by an old $p(x)$ in right-hand side. This new $p(x)$ exhibits a better estimate on capacity-achieving distribution. It has been proven that

$$p^t(x)\to p^*(x), I^t(X;Y)\to C. \quad (51)$$

when the recursive number $t$ goes to infinite [24, 25].

### B. Modified BA algorithm under multiple constraints

In a real channel, multiple constraints may be specified simultaneously. As the most common cases, PPC and APC are discussed extensively.

In the system under PPC, the object of constraint is the peak power independent of the input distribution. It is easy to utilize the BA algorithm to solve the capacity-achieving distribution by fixing the mass point location when the support constellation size is constant (specific modulation formats, such as PAM-2, PAM-4). Furthermore, the minimum capacity-achieving input cardinality is also extensively explored [26-29] in a PPC system, since the PS technique improves the energy efficiency at the expense of the increased pear-to-average power ratio (PAPR) and constellation expansion ratio [20]. In this paper, we utilize the DAB algorithm [30, 31] proposed in molecular communication system to find the capacity-achieving distribution with minimum cardinality for the IM-DD system under PPC. The heart of the application of DAB algorithm in the IM-DD system is based on the following facts:

1. The channel capacity is upper bounded by [32]

$$C=\max_x D[p(y\,|\,x)\,\|\,p(y)], x\in\chi=[x_1,x_2,...,x_M]. \quad (52)$$

---

**Algorithm 1** DAB algorithm for IM-DD system under PPC

**Initialization:** Select a mass points location vector $\chi^{(0)}=[x_1,x_2,...,x_N]$ in increasing order. We choose $N=2$ and a specific low PSNR $\eta$ for initialization. Select a tolerance $\epsilon$, which determines the maximum acceptable distance between the lower bound and the upper bound of capacity. The peak power is set to 1.

**Iterations:** Solve for the capacity-achieving input distribution $p^*(x)$, location $\chi^*$, and the capacity $C$
  1) Utilize the BA algorithm to obtain the optimal distribution $p^{(n)}(x)$ and lower bound of capacity $I^{(n)}$ for a given mass location vector $\chi^{(n)}$
  2) Compute the capacity upper bound with $p^{(n)}(x)$ and $\chi^{(n)}$

$$I_{up}^{(n)}=\max_{x\in\chi^{(n)}}D[p(y\,|\,x)\,\|\,p(y)].$$

  and $x_{max}^{(n)}\in[0\ 1]$ maximizes the upper bound

$$x_{max}^{(n)}=\arg\max_{x\in[0\ 1]}D[p(y\,|\,x)\,\|\,p(y)].$$

  if $I_{up}^{(n)}-I^{(n)}<\varepsilon$, $p^*(x)=p^{(n)}(x)$, $\chi^*=\chi^{(n)}$, $C=D_{max}^{(n)}$, $\eta=\eta+\Delta\eta$,
  otherwise continue.
  3) if $x_{max}^{(n)}=0.5$ or $x_{max}^{(n)}\in\chi^{(n)}$, $N^{(n+1)}=N^{(n)}+1$, otherwise $N^{(n+1)}=N^{(n)}$
  4) Optimize the location of mass points according to the solved $x_{max}^{(n)}$, and go to 1)

---

**Algorithm 2** Modified BA algorithm for IM-DD system under APC

**Initialization:** Select a constellation scaling vector $L=l_{min}:\Delta l:l_{max}$ and basic mass points with location $\chi_b=[x_1,x_2,...,x_M]$ and uniform distribution. Select an average power $P_{ave}$.
**Iteration:**
1) For a fixed constellation scaling factor $l$, utilize the BA algorithm to compute the maximum MI achieving distribution $p_l(x)$ and $I_l$ with

$$\chi_l=l\cdot\chi_b$$

2) Sweep $l$, perform 1) to find the optimal input distribution and MI for each $l$
3) Find the system capacity $C$ and the capacity achieving distribution $p^*(x)$

$$l^*=\arg\max_l I_l$$

$$C=I_{l^*}$$

$$p^*(x)=p_{l^*}$$

4) For another SNR, go to 1.

---

2. If the discrete input distribution $p(x)=[p_{x_1},p_{x_2},...,p_{x_M}]$ and the mass point location $\chi=[x_1,x_2,...,x_M]$ are optimal. For

$$x^*=\arg\max_x(D[p(y\,|\,x)\,\|\,p(y)]), x\in[0\ P_{peak}]. \quad (53)$$

The $x^*$ must be an element of $\chi$ [32].

3. If the input distribution $p(x)$ is optimal, it must be symmetric about $P_{peak}/2$ [26, 33].

4. The number of mass point increases monotonically and by at most one as the PPC is relaxed [26, 32, 33]. The new added mass point must be at $P_{peak}/2$ or the mass point $P_{peak}/2$ split into two mass points.

Detailed DAB algorithm is described in Algorithm 1.

For initialization, a mass point location vector $\chi^{(0)}=[x_1,x_2,...,x_N]$ is selected. As on-off-keying (OOK) is the optimal modulation format for extremely low PSNR regions in the PPC system, we set $N^{(0)}=2$, and start iteration in a low PSNR case with $PSNR=\eta$. In addition, a tolerance $\epsilon$ is utilized for the accuracy control of the final capacity, and the peak power is set to 1.

1) *Solve for the capacity and optimal distribution according to BA algorithm*:

We utilize the conventional BA algorithm to find the optimal distribution $p^{(n)}(x)$ and the low bound of the capacity $I^{(n)}$ based on the current mass point location $\chi^{(n)}$.

2) *Check for convergence*:

At the same time, the upper bound of the capacity can be obtained by

$$D_{max}^{(n)}=\max_{x\in\chi^{(n)}}D[p(y\,|\,x)\,\|\,p(y)]. \quad (54)$$

and the $x\in[0\ 1]$ maximizes the upper bound is numerically solved

$$x_{max}^{(n)}=\arg\max_{x\in[0\ 1]}D[p(y\,|\,x)\,\|\,p(y)]. \quad (55)$$

If the difference between the upper and lower bound is below the preset threshold $\epsilon$, the capacity, capacity-achieving distribution, and optimal mass point location are obtained as

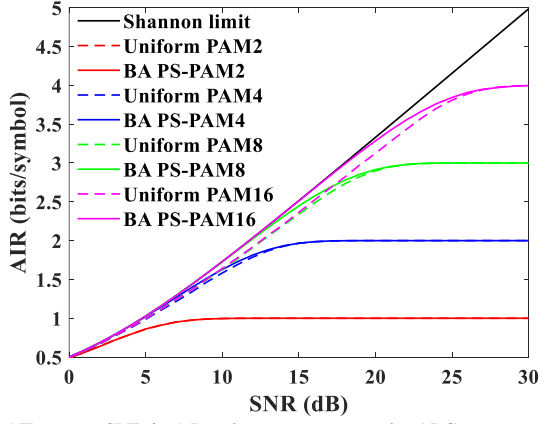

Fig. 4. AIR versus SNR in 1-D coherent systems under APC.

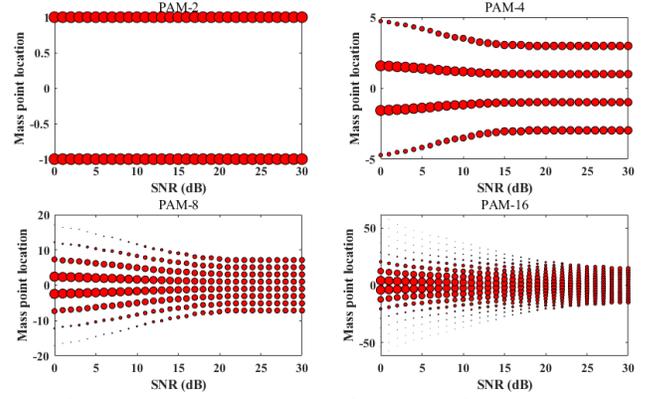

Fig. 5. Capacity achieving input PMFs for $|\chi|$=2, 4, 8, 16 in 1-D coherent systems. Size of the particle indicates probability.

$$p^*(x) = p^{(n)}(x) \ , \ \chi^* = \chi^{(n)} \ , \ C = D_{\max}^{(n)} \quad (56)$$

then go to 1) for another PSNR $\eta = \eta + \Delta\eta$. Otherwise go to 3)

3) *Add a new mass point if need:*

The relationship between solved $x_{\max}^{(n)}$ and current mass point location $\chi^{(n)}$ determines whether a new mass point needs to be added. If $x_{\max}^{(n)} = 0.5$ or $x_{\max}^{(n)} \in \chi^{(n)}$, a new mass point is added at 0.5, $N^{(n+1)} = N^{(n)} + 1$. Otherwise, $N^{(n+1)} = N^{(n)}$.

4) *Optimize the current mass point location*

If $x_{\max}^{(n)} < 0.5$ and $x_{\max}^{(n)} \in (x_j \ x_{j+1})$, we replace the $x_{j+1}$ in mass point location vector $\chi^{(n+1)}$ by $x_{\max}^{(n)}$. Due to the symmetric of the mass point, the $x_{N^{(n+1)}-j}$ should be replaced by $1 - x_{\max}^{(n)}$. If $x_{\max}^{(n)} > 0.5$ and $x_{\max}^{(n)} \in (x_j \ x_{j+1})$, we replace the $x_j$ by $x_{\max}^{(n)}$, and $x_{N^{(n+1)}+1-j}$ should be replaced by $1 - x_{\max}^{(n)}$. Thus, a new mass point location vector $\chi^{(n+1)}$ is generated. Then go to 1).

In an APC IM-DD system with fixed support constellation size, the average power is described by the first-moment of the signal, and the cost function can be further extended to

$$J = I(X;Y) + \lambda\left[\sum_\chi p(x) - 1\right] + \mu\left\{\sum_\chi\left[p(x) \cdot l \cdot x\right] - P_{ave}\right\}. \quad (57)$$

$$I(X;Y) = \sum_X \sum_Y p(x) \cdot p(y \mid x) \cdot \log_2\left[\frac{p(x \mid y)}{p(x)}\right]. \quad (58)$$

Hence,

$$p(x) = \frac{2^{\mu \cdot l \cdot x} \cdot p(x) \cdot 2^{D\left[p(y|x)\|p(y)\right]}}{\sum_\chi\left\{2^{\mu \cdot l \cdot x} \cdot p(x) \cdot 2^{D\left[p(y|x)\|p(y)\right]}\right\}}. \quad (59)$$

We rewrite the Eq. (59) to reflect the recursive nature

$$p^{t+1}(x) = \frac{2^{\mu \cdot l \cdot x} \cdot p^t(x) \cdot 2^{D\left[p(y|x)\|p(y)\right]}}{\sum_\chi\left\{2^{\mu \cdot l \cdot x} \cdot p^t(x) \cdot 2^{D\left[p(y|x)\|p(y)\right]}\right\}}. \quad (60)$$

It can be found that the updated $p(x)$ is related to the Lagrange multiplier $\mu$ and constellation scaling factor $l$. In fact, the two factors show a one-to-one mapping relationship due to the constraint of average power. For a given $l$, $\mu$ can be numerically solved by

$$\sum_X p^{t+1}(x) \cdot l \cdot x = P_{ave}. \quad (62)$$

Namely,

$$\sum_\chi\left\{2^{\mu \cdot l \cdot x} \cdot 2^{D\left[p(y|x)\|p(y)\right]} \cdot (l \cdot x - P_{ave})\right\} = 0. \quad (63)$$

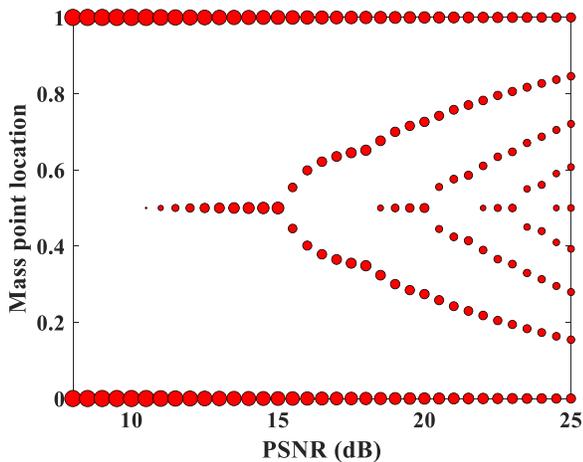

Fig. 6. Capacity achieving distribution with minimum cardinality for PPC IM-DD systems. Size of the particle indicates probability.

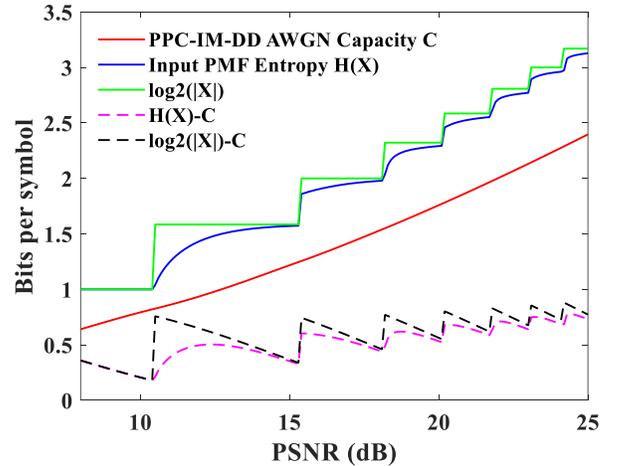

Fig. 7. Capacity, input entropy, and Log minimum input cardinality for PPC IM-DD systems.

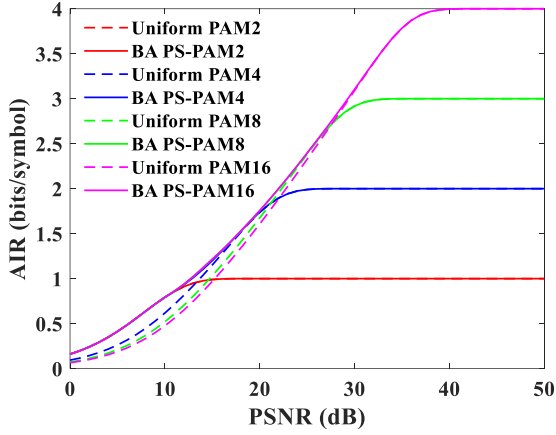

Fig. 8. AIR versus PSNR in PPC IM-DD systems with fixed constellation cardinality.

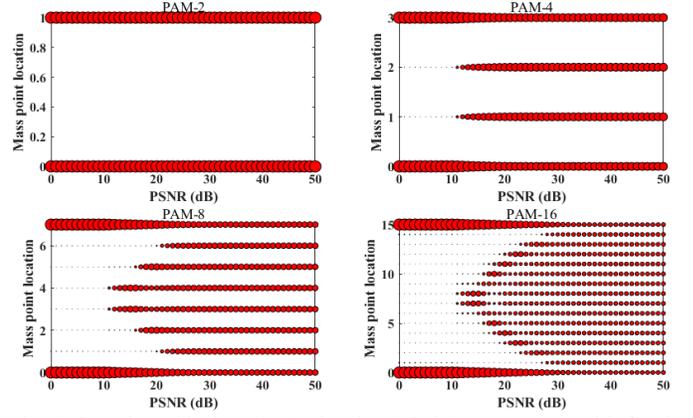

Fig. 9 Capacity-achieving distribution in PPC IM-DD systems with fixed constellation cardinality. Size of the particle indicates probability.

Thus, the capacity-achieving distribution $p_l(x)$ and the mutual information $I_l$ under a fixed constellation scaling factor $l$ can be numerically calculated by Eq. (60) and Eq. (58). Finally, the $l$ is swept to find the optimal distribution and maximum MI with the $l^*$.

$$p^*(x) = p_{l^*}(x), C = I_{l^*}. \tag{63}$$

Consequently, the modified BA algorithm for the capacity-achieving distribution searching in the IM-DD system under APC can be concluded as Algorithm 2:

*Step 1*: Give a SNR and constellation scaling factor $l$, utilizing the BA algorithm to find the MI achieving distribution $p_l(x)$ and MI $I_l$

*Step 2*: Sweep $l$ in a given range and step, perform step 1 to find the optimal input distribution and MI for each $l$.

*Step 3*: Find the system capacity $C = \max_l I_l$ and the capacity achieving distribution $p_{l^*}$.

*Step 4*: $SNR = SNR + \Delta SNR$, go to step 1.

## IV. RESULTS

This section utilizes the mentioned BA algorithms to evaluate the capacity and capacity-achieving distribution for different system models. For comparison, we present the results of 1-D coherent system (bipolar PAM) first, in which the capacity is also obtained by BA algorithm. Then, the IM-DD system under PPC and APC are discussed in detail.

### A. Coherent systems

For long-haul coherent fiber-optic systems, we utilize the Algorithm 2 to calculate the capacity and capacity-achieving distribution with fixed mass point cardinality, since the optical amplifier invalidates the PPC and imposes the APC. In this system, the average power of signal is represented by the second moment as

$$\sum_\chi p(x) \cdot \| l \cdot x \|^2 = P_{ave}. \tag{65}$$

where $\chi$ is the bipolar PAM symbol without nonnegative constraint. Fig. 4 shows the achieved information rate (AIR) versus SNR for some specific modulation formats including PAM2, PAM4, PAM8, and PAM16. The solid line depicts the AIR of the capacity-achieving distribution obtained by the iterative BA algorithm, and the dash line shows the AIR of the uniform PAM signal computed by Monte Carlo simulation. It can be found that there is no shaping gain for PAM2 format as the two symbols exhibit the same power. As the modulation format increases, so does the shaping gain. The detailed capacity-achieving distributions for those specific modulation formats are given in Fig. 5. Under different SNRs, the location of the mass point can be read from the y-axis, and the probability of the mass point is indicated by the size of the particles. It can be found that the mass point with higher power shows a lower probability, actually, which satisfies a discrete Gaussian distribution. If the SNR is high enough, the capacity-achieving distribution turns to a uniform distribution for a fixed modulation format.

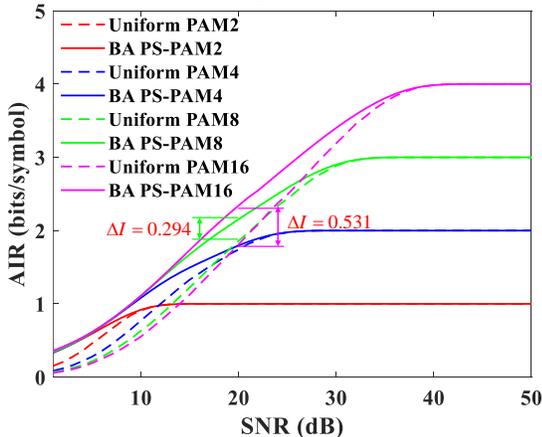

Fig. 10. AIR versus SNR in APC IM-DD systems with fixed constellation cardinality.

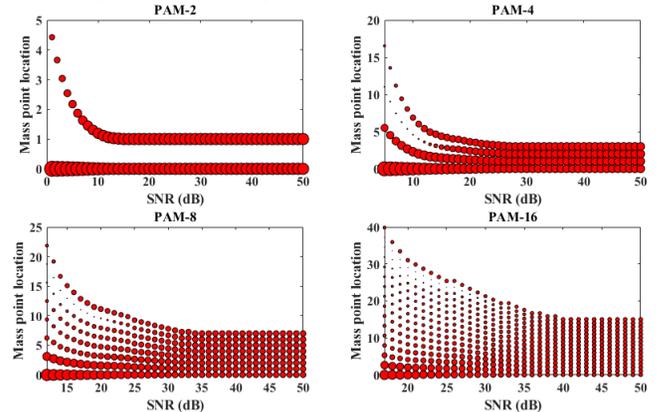

Fig. 11 Capacity-achieving distribution in APC IM-DD systems with fixed constellation cardinality. Size of the particle indicates probability.

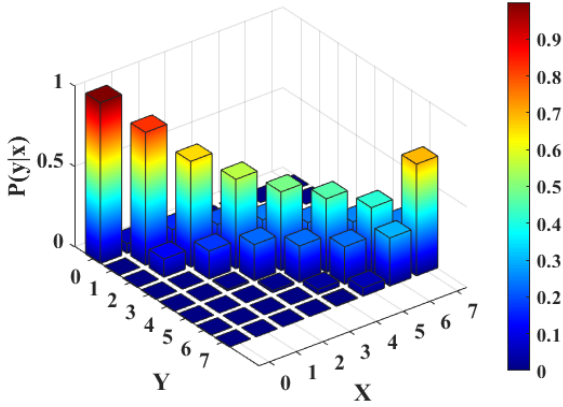

Fig. 12. Discrete $p(y=k|x=j)$ of PAM8 format under 20dB SNR.

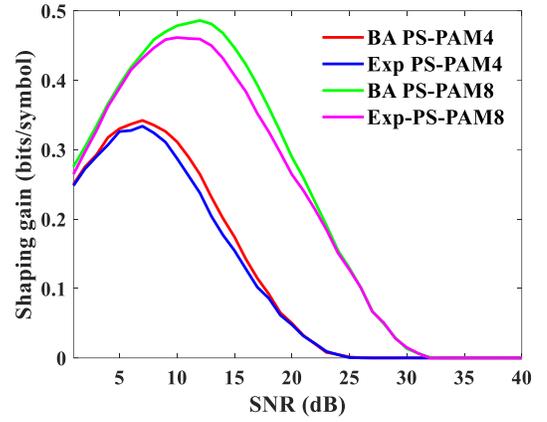

Fig. 13. Shaping gain versus SNR of discrete exponential distribution and BA optimized distribution.

### B. IM-DD systems under PPC

For IM-DD systems without optical amplifier, the system is limited by the peak power. For such a system, the PSNR is a proper channel metric. Before evaluating the capacity-achieving distribution for a fixed modulation format, we utilize the DAB algorithm, namely Algorithm 1, to compute the capacity-achieving distribution with minimum cardinality in a wide range of PSNR, and the results are shown in Fig. 6. In the low PSNR regions, uniform PAM2 is always the optimal solution. As the PSNR increases, a mass point with low probability appears in the middle of the two symbols, which means that non-uniform PAM3 with a high capacity beats the PAM2. As the PSNR continues to increase, either a new mass point appears in the middle location (0.5 in this paper) when the current cardinality of the input symbol is even, or the symbol 0.5 is split into two new mass points when the current cardinality of the symbol is odd. The corresponding capacity $C$, optimal input entropy $H(X)$, and minimum input cardinality $\log_2(|X|)$ are shown in Fig. 7. As the PSNR increases, the required resources, including input entropy and cardinality of input symbol, to achieve a higher capacity grow in a predictable way. The difference between the input entropy $H(X)$ and system capacity $C$ is also given out in Fig. 7 with rose red dashed line, which indicates the information loss between the transmitter and receiver side. Furthermore, the difference between the required minimum cardinality of the input symbol $\log_2(|X|)$ and capacity $C$ is shown in Fig. 7 with black dash line. A maximum cardinality excess of approximately 0.8 bits above the capacity is observed in the PPC IM-DD system.

The results shown Fig. 6 enabled by DAB algorithm can be considered as the combination of 1-D geometric shaping and PS. In a practical system, the cardinality and the location of the input signal are always constrained in a PPC system. Hence, we investigate the capacity and capacity-achieving distribution in PPC IM-DD system with a fixed constellation size and symbol location, and the results of AIR versus PSNR are given in Fig. 8. It can be found that in such a system, the shaping gain can only be obtained in the low PSNR regions, which is quite different from the coherent system. In Fig. 9, we depict the capacity-achieving distribution for several specific modulation formats. It can be observed that the capacity-achieving distribution is symmetric, which is consistent well with the results in [26, 32, 33]. For PAM4 format, it shows a cup shape, but for higher modulation formats, it shows an irregular shape which has not been reported. As the PSNR increases, the capacity-achieving distribution tends to the uniform distribution for a fixed modulation format. Although a small shaping gain is observed in this kind of system, the PS technique can also be utilized to achieve the rate adaptability by introducing the PAS structure, since the optimal distribution is symmetric.

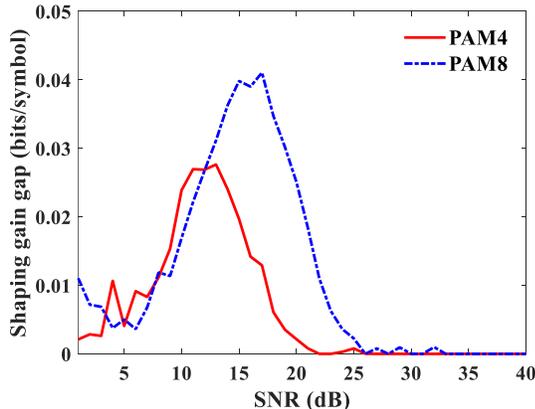

Fig. 14. Shaping gain gap between BA optimized distribution and discrete exponential distribution of PAM4 and PAM8 modulation formats.

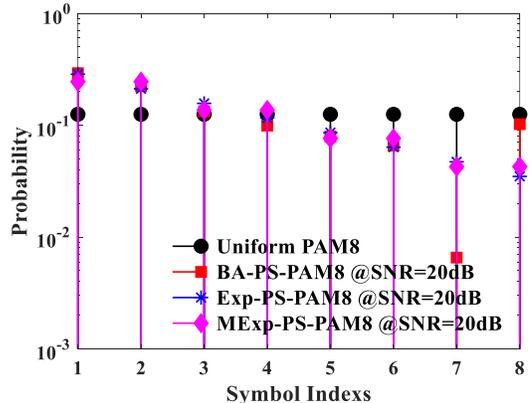

Fig. 15. Different input distributions of PAM8 format at 20dB SNR.

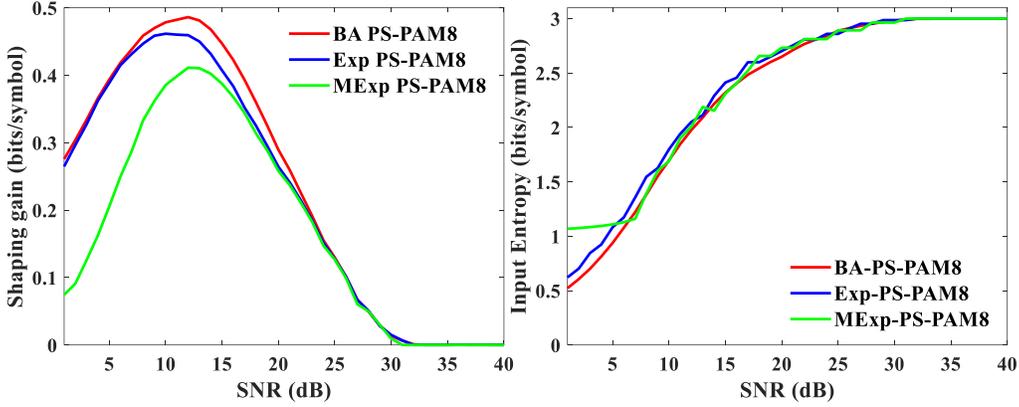

Fig. 16. (a) Shaping gain versus SNR for PAM8 format. (b) Optimal input entropy for PAM8 format with different distributions.

### C. IM-DD systems under APC

#### 1) Theoretical simulation

For some specific applications such as intra-DCI, the optical amplifier is required to boost the signal to yield a desired launch power or compensate the transmission loss at the receiver, which invalidates the PPC and imposes the APC. In this case, we utilize the Algorithm 2 to calculate the capacity and capacity-achieving distribution with a fixed constellation cardinality. Fig. 10 shows the achieved capacity of several modulation formats in a wide range of SNR, and the corresponding capacity-achieving distributions are depicted in Fig. 11. It is obvious that the optimal distribution is quite different from the coherent system and PPC IM-DD system. In the low SNR regions, the uniform PAM2 is no longer the best choice. Shaping gain of the PAM2 modulation format can also be obtained by reducing the probability of the symbol with higher energy. For higher modulation formats, the shaping gain is more evident. When the SNR is 20dB, the shaping gain of PAM8 and PAM16 formats can realize 0.294 bits/symbol and 0.531 bits/symbol, respectively. As shown in Fig. 11, the capacity-achieving distribution is irregularly shaped, especially for the symbol with highest energy. Intuitively, the symbol with higher energy will be assigned to a lower probability in the APC system to obtain the power efficiency. But in Fig. 11, the symbol with the highest energy exhibits a higher probability than the adjacent symbol (lower energy) in high modulation formats. This phenomenon can be explained by the Fig. 12, where we give out the discrete conditional probability function $p(y=k|x=j)$ for PAM8 modulation format with the model described in Eq. (34) under 20dB SNR. The diagonal elements of the discrete conditional probability function reflect the reliability of the channel for different transmitted symbols. The highest energy symbol "7" exhibits a higher transmission reliability than the symbol "6". Thus, as a numerical iterative optimization method, the BA algorithm assigns a higher probability for the symbol "7".

For the APC IM-DD system, we conclude in Section II that the exponential distribution is the local optimal solution for system with infinite SNR. However, the shaping gain between the BA optimized distribution and discrete exponential distribution cannot be solved analytically. Here, we numerically compute the capacity of the BA optimized distribution and discrete exponential distribution, and the shaping gain for PAM4 and PAM8 modulation formats are given out in Fig. 13. It can be found that the shaping gain difference between the BA optimized distribution and discrete exponential distribution only appears at the moderate SNR regions. We depict the detailed shaping gain gap of PAM4 and PAM8 formats in Fig. 14. The maximum gap is lower than 0.05 bits/symbol for PAM8, which is observed at around 16dB SNR.

#### 2) Practical discussion

Although a significant shaping gain in terms of the power efficiency can be obtained in APC IM-DD systems, the capacity-achieving distribution shows an asymmetrical characteristic, which is different from the long-haul coherent system and PPC IM-DD system. This indicates that it is impossible to utilize the PAS structure to realize the rate adaptation in the ACP IM-DD system with the solved capacity-achieving distribution. Based on the observation of negligible shaping gain between the discrete exponential distribution and BA optimized distribution, a modified exponential distribution with pairwise probability can be considered as a candidate practical solution for the application of PS technique in the ACP IM-DD system. Fig. 15 shows the different input distributions of PAM8 symbol. In modified exponential distribution, every two symbols exhibit the same probability, which can be realized by a two-step processing: 1) generate the exponentially distributed PAM4 symbols utilizing CCDM, and label the PAM4 symbol with two bits by Gray mapping which can be considered as the first two bits of the PAM8 symbol; 2) the last bit of the PAM8 symbol can be added by the uniform bit (check bits or addition signal bits). For any other modulation formats, it can also be generated by this way.

By introducing the modified exponential distribution, the rate adaptability of the system can be well guaranteed. However, the shaping gain of this bounded distribution should be re-assessed. Fig. 16 (a) shows the shaping gain evaluation of the PAM8 format with different distributions. It can be found that the significant shaping gain loss of the modified exponential distribution appears at the low SNR regions. These low SNR regions are out of our interest due to the ultra-low AIR. Fig. 16(b) shows the optimal input entropy of these distributions. The input entropy of the modified exponential distribution is lower bounded by 1 due to the fact of pairwise shape. This is the main reason for the loss of shaping gain in low SNR regions.

## V. CONCLUSION

In this work, a theoretical evaluation of the capacity and capacity-achieving distribution for IM-DD fiber-optic channels is investigated. In order to exactly discuss this issue, two system models are established: 1) PPC IM-DD systems, which are suitable for the system without optical amplifier; 2) APC IM-DD systems, which are related to the system with optical amplifier. Different from the long-haul coherent system, the analytic solution of capacity-achieving distribution cannot be directly found in the aforementioned two IM-DD systems. Alternately, an iterative method, BA algorithm, is utilized to numerically compute the capacity and capacity-achieving distribution. For the PPC IM-DD system, a dynamic-assignment BA algorithm is applied to find the minimum cardinality input capacity-achieving distribution. It is observed that the maximum difference between the minimum input cardinality and capacity is around 0.8 bits. For a fixed support constellation size, the capacity and capacity-achieving distribution is easily solved by the conventional BA algorithm in PPC IM-DD system. Although a small shaping is observed in this case, the PAS technique can also be used to introduce rate adaptation to the system by adjusting the shaping and FEC overheads, since the capacity-achieving distribution is symmetric. In the APC IM-DD system, a modified BA algorithm is investigated to calculate the capacity and capacity-achieving distribution, and a significant shaping gain is observed. For PAM8 and PAM16 modulation formats, 0.294 bits/symbol and 0.531 bits/symbol shaping gain can be obtained at 20dB SNR. However, the optimal distribution shows an asymmetrical characteristic, which cannot be directly utilized in the PAS structure for the rate adaption realization. Thus, the practical application of PS technique in this situation is further discussed.

## APPENDIX

Considering utilizing the Lagrange multiplier method to solve the maximum entropy achieving distribution for PPC and APC IM-DD systems. For PPC IM-DD systems, the cost function can be expressed as Eq. (33)

$$J = H(X) + \lambda \left[ \sum_{\chi} p(x) - 1 \right] + \mu \left[ \max(x) - P_{peak} \right]. \quad (65)$$

$$\frac{\partial J}{\partial p(x)} = -\log_2 [p(x)] - \frac{1}{\ln 2} + \lambda. \quad (66)$$

$$p(x) = 2^{\lambda - \frac{1}{\ln 2}}. \quad (67)$$

It can be found that the $p(x)$ is a constant value for different $x$, namely the solution is a uniform distribution. Normalizing the solved $p(x)$, it is expressed as

$$p(x) = \frac{1}{M}. \quad (68)$$

For APC IM-DD system, the cost function can be defined as Eq. (40)

$$J = H(X) + \lambda \left[ \sum_{\chi} p(x) - 1 \right] + \mu \left[ \sum_{\chi} [p(x) \cdot l \cdot x] - P_{ave} \right]. \quad (69)$$

$$\frac{\partial J}{\partial p(x)} = -\log_2 p(x) - \frac{1}{\ln 2} + \lambda + \mu \cdot l \cdot x. \quad (70)$$

$$p(x) = 2^{\lambda - \frac{1}{\ln 2} + \mu \cdot l \cdot x}. \quad (71)$$

For a fixed constellation scaling factor $l$, the Lagrange multipliers $\lambda$ and $\mu$ can be solved by the two constrains

$$\sum_{\chi} p(x) = 1. \quad (72)$$

$$\sum_{\chi} p(x) \cdot l \cdot x = P_{ave}. \quad (73)$$

Thus, the $p(x)$ expressed as Eq. (70) is the exponential distribution. In this case, the shaping gain can be calculated by

$$\Delta H = H(X_E) - H(X_U) \quad (74)$$

For uniform distribution in $[0 \; A]$, the PMF and average power are $1/A$ and $A/2$, respectively. The $H(X_U)$ is calculated by

$$H(X_U) = -\int_0^A p(x) \log_2 p(x) dx = \log_2 A. \quad (75)$$

As $P_{ave} = \dfrac{A}{2}$, the $H(X_U)$ can be further expressed as

$$H(X_U) = \log_2(2P_{ave}). \quad (76)$$

For a normal exponential distribution, the PMF is always given by

$$p(x) = \lambda \exp(-\lambda x), \lambda > 0. \quad (77)$$

The average power (first moment) and entropy can be expressed as

$$P_{ave} = \int_0^\infty p(x) \cdot x dx. \quad (78)$$

$$H(X_E) = -\int_0^\infty p(x) \log_2 p(x) dx. \quad (79)$$

Replacing the Eq. (77) into Eq. (78) and Eq. (79), the average power and entropy are $1/\lambda$ and $\log_2(e/\lambda)$, respectively. Then the $H(X)$ can be further expressed as

$$H(X_E) = \log_2 e P_{ave} \quad (80)$$

Thus, the shaping gain can be obtained

$$\Delta H = H(X_E) - H(X_U) = \log_2 \left( \frac{e}{2} \right). \quad (81)$$

Transforming the MI gain into power efficiency

$$\Delta P_{dB} = \frac{\partial P_{dB}}{\partial P} \cdot \frac{\partial P}{\partial H} \cdot \Delta H = 1.3326 dB. \quad (82)$$